\documentclass[review]{elsarticle}

\journal{Biomaterials}

\begin{document}

\begin{frontmatter}

\title{Probing Three-dimensional Collective Cancer Invasion with DIGME}

\author{Amani A. Alobaidi and Bo Sun\fnref{myfootnote}}
\address{110 Weniger Hall, Department of Physics, Oregon State
  University Corvallis, OR, United States}
\fntext[myfootnote]{correspondence send to {sunb@onid.orst.edu}}




\begin{abstract}
Multicellular migration and pattern formation play important roles in
developmental biology, cancer metastasis and wound healing. To
understand the collective cell dynamics in three dimensional
extracellular matrix (ECM), we have
developed a simple and mechanical-based strategy, Diskoid In Geometrically
Micropatterned ECM (DIGME). DIGME allows easy engineering of the shape
of 3-D tissue organoid, the mesoscale ECM heterogeneity, and
the fiber alignment of collagen-based ECM all at the same
time. We have employed DIGME to study the collective cancer invasion
and find that DIGME provides a powerful tool to probe three
dimensional dynamics of tumor organoid in patterned microenvironment.
\end{abstract}

\begin{keyword}
tumor organoid\sep extracellular matrix\sep collagen \sep collective invasion
\end{keyword}

\end{frontmatter}


\section*{Introduction}
Invasion in three-dimensional (3D) extracellular matrix (ECM) is an
important step in the lethal metastasis of tumors
\cite{Alsarraj2008}. Although extensive studies have elucidated
detailed mechanisms of single cell 3D motility
\cite{Brocker2000,grinnell2010,Yamada2012}, and cell-ECM interactions
\cite{Paluch2012,Yamada2013,Yamada2015}, 3D collective cancer invasion
is still poorly understood \cite{Gilmour2009,Segall2012}. Most studies
to date have focused on 2D collective cell migration. It has been
shown that cell-cell adhesion \cite{Gardel2011,Manning2015,Trepat2015}, exclusion
volume \cite{collective_glassy}, contact
inhibition \cite{Sun2011landscape,Levine2016}, cell-secreted chemical factors
\cite{chaserun}, and substrate-mediated mechanical forces
\cite{ma2013} coordinate the multicellular motility and pattern
formation of multiple cells in 2D. These results, however, have
limited applicability in 3D tumor progression. The topological
connectivity and porosity of 3D ECM allow cells to avoid touching one
another while migrating, thus mechanical signaling via direct
cell-cell contact is less important for collective motion in 3D
than in 2D. Similarly, chemical signaling in 3D suffers from
rapid dispersive dilution, thus the diffusion-mediated 3D
intercellular correlations are much weaker compared to the case in 2D
\cite{Sun2012collective}.

To probe the collective cell migration and morphogenesis, 2D cell
patterning and substrate engineering has provided much insights. For
instance, various types of wound healing assays have been developed to
explore the invasion of cancer cell colonies into extracellular voids
of pre-defined geometries
\cite{collective_wound,Trepat2014,Ladoux2015}. These assays typically
use soft-lithography fabricated stamps when seeding the cells, and
lift the stamps after the cells have adhered to the
substrate. Alternatively, geometric patterned cell adhesive and cell
repellent \cite{Ingber1994,Wang2011,Silberzan2014}, as well as
microfluidics channels \cite{Williams2011} have been employed to
restrict cell migration. By engineering the confining geometry of the
substrate, emergent multicellular dynamics, such as spontaneous
rotation in circular geometry \cite{Marchetti2016}, and directed
migration in ratchet geometry \cite{Grzybowski2009} have been
observed.

To probe the 3D collective cancer invasion, multicellular tumor
spheroid model has been widely employed
\cite{Sutherland1970,Sutherland1971,Schughart2010}. Tumor spheroids
are aggregates of cancer cells that preserve the native 3D cell-cell
contact, mimicking the configuration of solid
tumors\cite{Marie2015}. Multiple methods have been developed to grow
3D tumor spheroids, such as the hanging droplets \cite{Nielsen2007},
non-adhesive microplates \cite{Zhao2015}, and bi-phase liquid systems
\cite{Park2015}. However, these techniques can neither control the
geometry of the cell aggregates, nor have the capability of
engineering complex extracellular environment. Other methods, such as
3D tissue printing \cite{Bayley2013} and photo-sensitive hydrogel
\cite{West2006,Anseth2009,Anseth2011,Omenetto2015} are capable of 3D
cell-ECM patterning at the price of expensive equipment, non-native
ECM composition, or sophisticated sample preparation
\cite{Kanamori2016}. As an alternative, we have developed a low-cost, flexible strategy, Diskoid In Geometrically
Micropatterned ECM (DIGME). DIGME is mechanical-based, and is
compatible with a wide range of cells and ECM types. As we will
demonstrate below, DIGME combines the powers of 3D tumor organoids and
3D ECM patterning, allowing us to independently control the shape of
tumor organoids, microstructure and spatial heterogeneity of the ECM
all at the same time.

\section*{Materials and Methods}
\subsection*{mechanical setup}
The basic setup of DIGME consists of a x-y-z translational stage to hold
sample dish, and a rotational motor to mount the mold above the sample
stage. We have used parts from Thorlabs Inc. and TA instruments to assemble
prototypes of DIGME. When necessary, we have also placed DIGME setup
on an inverted microscope (Leica Microsystems) to help with alignment
and positioning the mold. See SI Fig. S1 for schematic design of the
DIGME setup.
\subsection*{prepare collagen gel}
High concentration collagen solution (10 mg/ml, Corning) is diluted and neutralized to
desirable concentrations with cell growth medium (see cell culture),
NaOH and 10X PBS, all purchased from Sigma-Aldrich. The neutralized,
ice cold solution is first poured into the sample dish mounted on the
DIGME setup where the mold is approximately 200 $\mu$m above the glass
bottom of the dish. After treated in the DIGME setup for 5-10 minutes
(with the mold statically immersed or rotatin in the geling solution),
we lift the mold out of the dish via the z-motor of the translational
stage. The collagen solution will continue the gelation process for
another 40 minutes. The molded gel is then immersed with fresh growth
medium and stored at 4 $^\circ$C for up to 2 days before adding cells.
\subsection*{cell culture and microscopy}
GFP-labeled MDA-MB-231 cells (Cell Biolabs Inc.) is maintained
according to the vendor's protocol. After embedding the cells, DIGME
devices are kept in tissue culture incubator except when taken out for
imaging. For confocal imaging, we use a Leica SPE microscope. 10X oil
immersion objective is used when confocal reflection imaging of
collagen fiber is needed. Otherwise, 4X air objective is used to image
the fluorescently labeled cells and fluorescent particles embedded in
the collagen matrix. The z-stacks of confocal imaging are taken with 2
$\mu$m z-steps. Confocal images are further processed in NIH ImageJ
and MATLAB.

\section*{Results}

\begin{figure}[t]
\centering
  \includegraphics[width=0.75 \columnwidth]{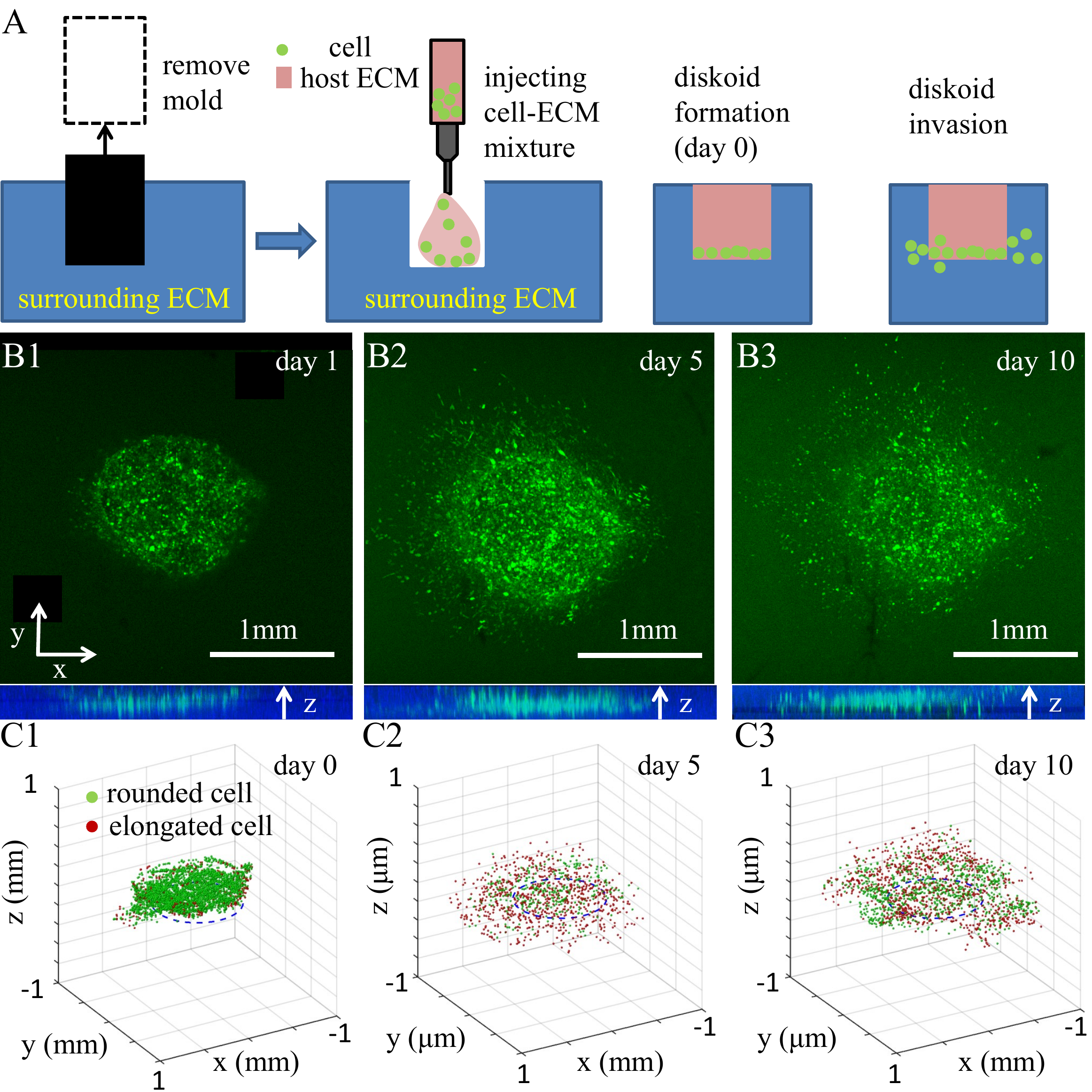}
  \caption{Preparation and collective migration of a circular
    DIGME. (A) Schematics showing the steps of forming a DIGME
    device. (B1-B3) Top views of a diskoid in 3D ECM. GFP-labeled
    MDA-MB-231 cells are cultured in DIGME device and confocal imaging
  are performed at day 1, 5 and 10. Bottom insets show the
  corresponding side views. (C1-C3) Manually identified 3D cell centers
  and morphological phenotypes corresponding to B1-B3. Green: rounded
  cells with aspect ratio less than 2. Red: elongated cells with
  aspect ratio greater than 2.}
  \label{fig1}
\end{figure}
To demonstrate the working principles and biocompatibility of DIGME,
we first formed a cylindrical MDA-MB-231 tumor diskoid in 3D type-I
collagen gel. Briefly, a stainless steel needle is used to mold the
collagen gel (surrounding ECM) with a cylindrical well. The well is
then filled with neutralized cell-collagen solution. After cells
quickly sediment down to the well bottom (within 1 minute), collagen
solution continues to polymerize and eventually forms the host ECM
that covers the cell aggregate -- a diskoid -- on the bottom of the
well (Fig. \ref{fig1}A). Within 24 hours of incubation, cells start to
invade into the surrounding ECM. Fig. \ref{fig1}B1-B3 demonstrate the
top views (x-y plane) and side views (x-z plane) of a DIGME sample at
day 1, 5 and 10. Notice that although the surrounding ECM and the host
ECM are polymerized at the same temperature (21 C$^\circ$) and have
the same concentration (1.5 mg/ml), the invasion in the radial
direction is much more pronounced compared with the spreading in the z
direction. The biased migration direction is presumably a collective
phenomena due to the cell-cell interactions \cite{Sun20133D,Wirtz2015}. 

DIGME allows continuous confocal imaging at the single cell level,
therefore we can track the morphological profiles of the diskoid over
time. Cells in the diskoid generally exhibit two distinct
morphologies: elongated cells are typically fast moving and strongly
contracting, while rounded cells migrate with short persistence and
exert only weak traction forces
\cite{Wirtz2013,Sun2016reflectance}. Empirically, we distinguish
elongated and rounded cells based on the cell aspect ratio with a
threshold value of 2. We have manually located the center of each cell
and have classified each cell into elongated or rounded phenotypes as
shown in Fig. \ref{fig1}C1-C3.

\begin{figure}[t]
\centering
  \includegraphics[width=0.75 \columnwidth]{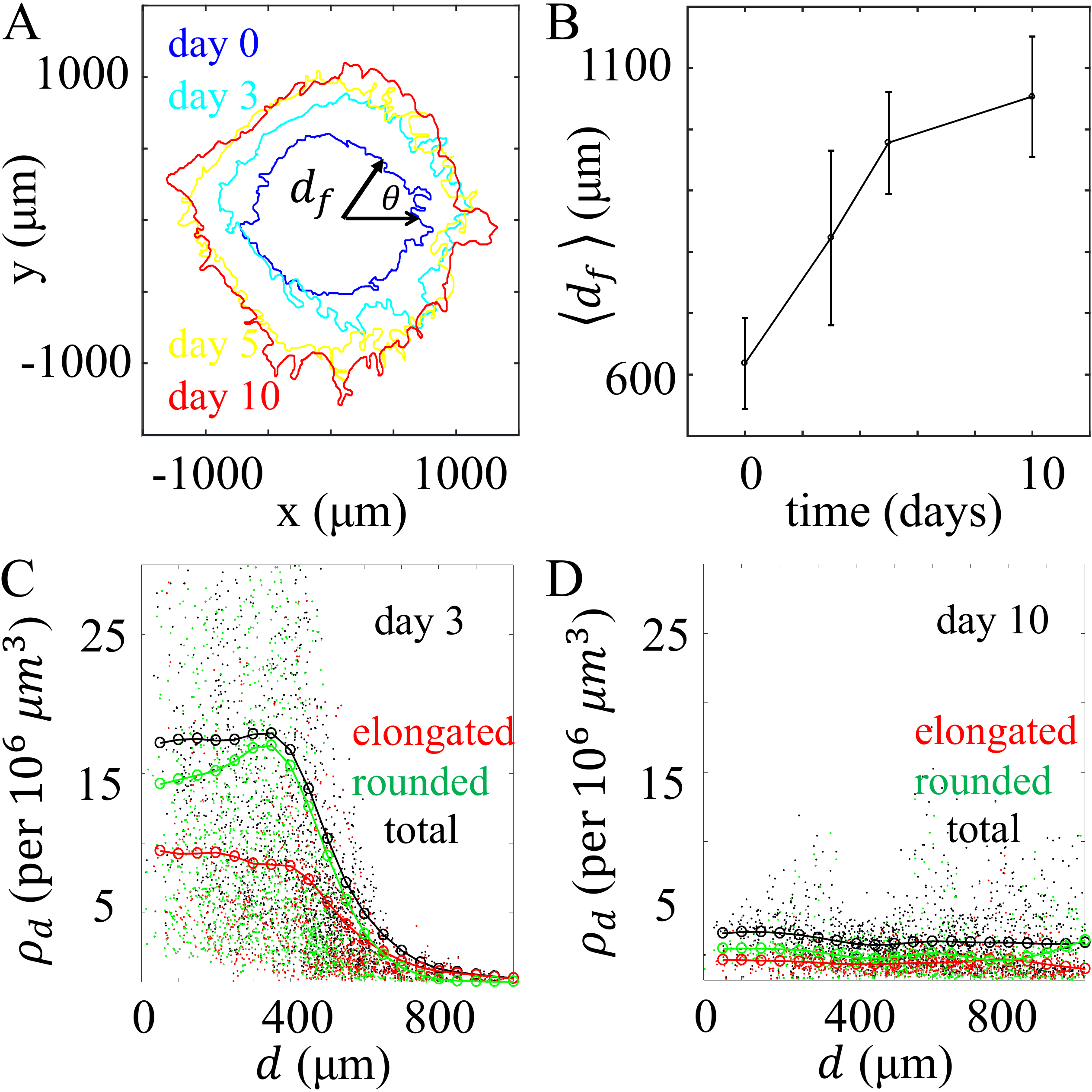}
  \caption{The invasion profile of a DIGME device. (A) The invasion
    fronts at day 0, 3, 5, and 10 of the same sample described in
    Fig. \ref{fig1}. (B) The means and standard deviations of the
    invasion distance obtained by averaging $d_f(\theta)$ over the
    polar angle $\theta$ of (A). (C-D) Scattered plots: the local cell
  density by counting only the elongated cells (red), or only the
  rounded cells (green), or all cells (black). Lines: Running average of the
  scattered data points with a Gaussian kernel. (C) shows the results
  at day 3, (D) shows the results at day 10.}
  \label{fig2}
\end{figure}
In order to quantify the diskoid invasion profile, we have
located the invasion fronts by projecting the confocal images onto the x-y
plane. The invasion front can be described as $d_f(\theta,t)$, where
$d_f$ measures the radial distance from the center of the well, $\theta$ is
the polar angle, and $t$ is the time of diskoid
invasion. Fig. \ref{fig2}A show the invasion fronts at day 0, 3, 5,
and 10. After averaging over the polar angle $\theta$, Fig. \ref{fig2}B show
the means and standard deviations of $d_f$. The invasion front grows
sublinearly with time, which is consistent with the observation on
tumor spheroid.

In order to quantify the morphological evolution of the diskoid, we
have calculated the elongated, rounded and full cell density using
k-nearest neighbors of each cell. Briefly, for each cell $i$
at location $\mathbf{r}_i = [x_i, y_i, z_i]$, we find the minimal
sphere centered at $\mathbf{r}_i$ with radius $r_m$ that encose
exactly k cells. The cell density at $\mathbf{r}_i$ is approximated to
be $\rho_d(\mathbf{r}_i) = \frac{3k}{4\pi r_m^3}$. For simplicity, we
have chosen $k=10$. Fig. \ref{fig2}C-D show the cell density at
varying invasion depth $\rho_d(d)$, where
$d=\sqrt{x^2+y^2}$. At day 3, only a small number of cells have
migrated far from the seeding radius (original diskoid-ECM interface) at $a=$370
$\mu$m, and these cells are mostly elongated. Close to the center,
cell density is approximately constant for $d\leq$ 300 $\mu$m. As
invasion proceeds, the region of constant cell density expands. At the
same time, cell density in this region decreases because the cell
proliferation is slow compared to the migration-induced
dilution. After invading the surrounding ECM for 10 days, both
elongated and rounded cells are uniformly distributed for $d\leq$ 1
mm, and the cell density has dropped by more than four folds.
\begin{figure}[t]
\centering
  \includegraphics[width=0.75 \columnwidth]{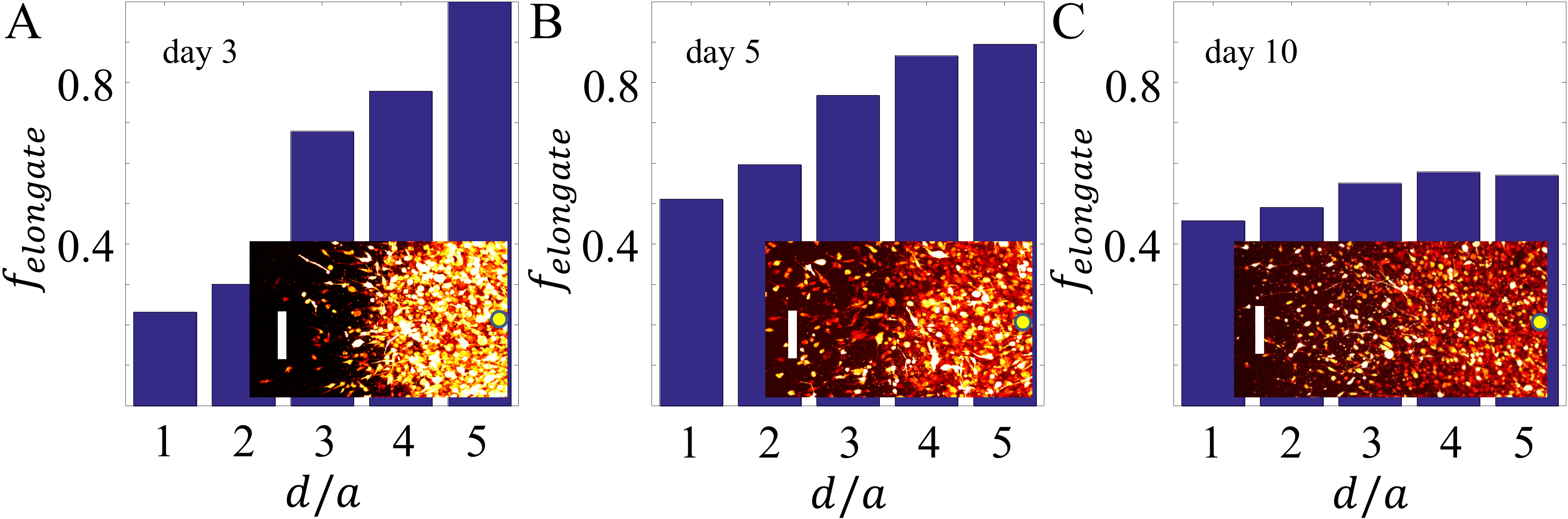}
  \caption{The spatial-temporal profiles of cell morphology. (A-C)
    Fraction of elongated cells $f_{elongate}$ at varying distances
    from the center. Here the distance is normalized by the seeding
    radius $a=$ 370 $\mu$m of
    the same diskoid described in Fig. \ref{fig1}. Insets: a section
    of the top view invading diskoid taken at day 3, 5 and 10. Scale bars of the
    insets: 200 $\mu$m.}
  \label{fig3}
\end{figure}
To further quantify the morphological distribution, we have normalized
the invasion depth $d$ with respect to the seeding radius $a$ of the
diskoid and have calculated the fraction of elongated cells
$f_{elongate}$ in different regions of $d/a$. As shown in
Fig. \ref{fig3}A, at day 3 $f_{elongate}$ increases rapidly at greater
radial distance, consistent with the fact that elongated cells are
scout cells during collective invasion. The positive correlation
between $f_{elongate}$ and $d/a$ gradually diminishes over time
(Fig. \ref{fig3}B). At day 10, a half-half mixture of elongated and
rounded cells are found in all regions of the sample. To
quantitatively account for these observations, we have developed a
simple model based on persistent random walks. We assume the cells
stochastically transform between elongated and rounded phenotypes at a
rate of $\tau_{trans}$, and the two phenotypes have distinct migration
persistent time $\tau_{el}$ and $\tau_{rd}$. As elaborated in the SI,
the model agrees well with the results of Fig. \ref{fig2}
and Fig. \ref{fig3}.

\begin{figure}[t]
\centering
  \includegraphics[width=0.75 \columnwidth]{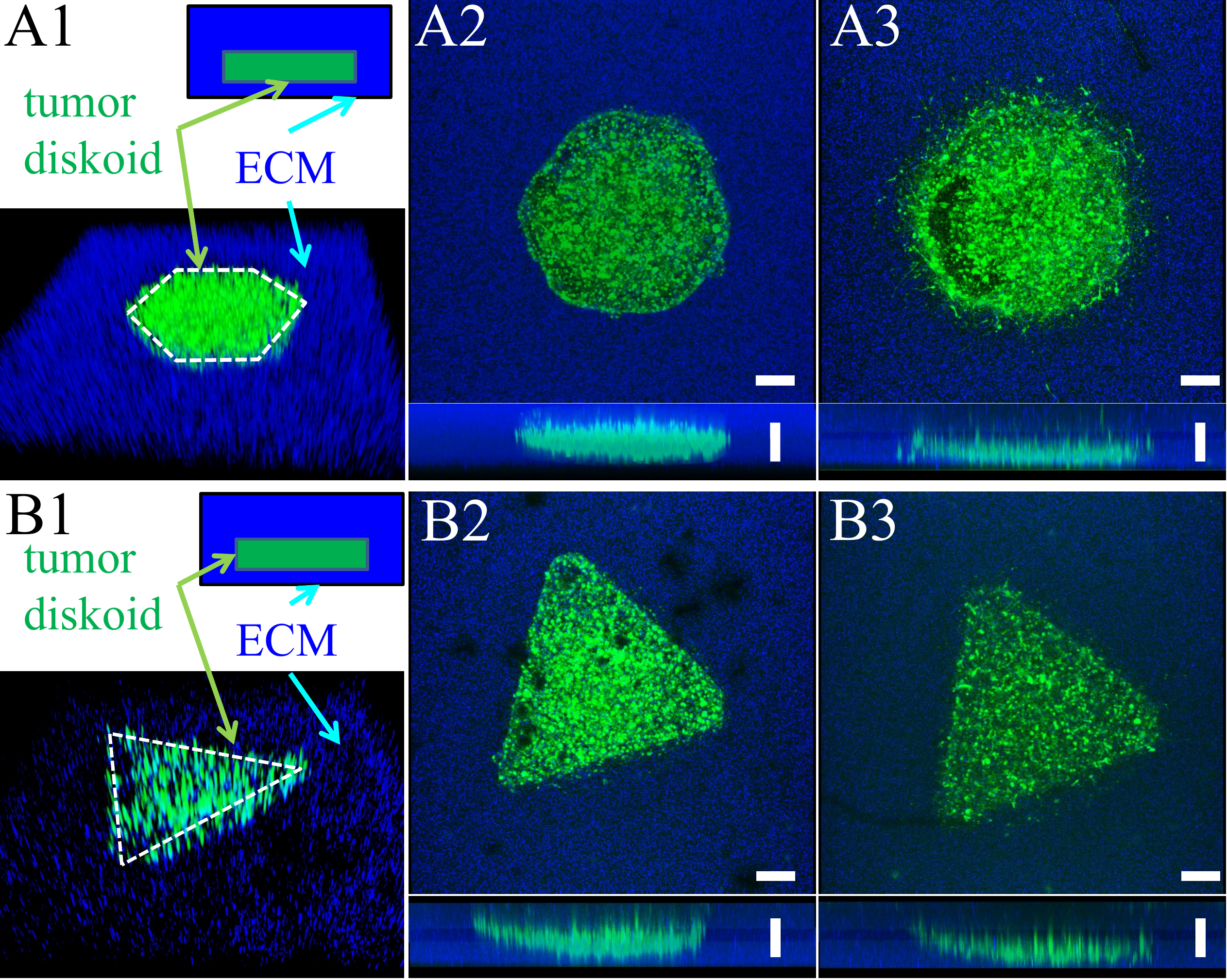}
  \caption{The hexagonal and triangular diskoid in DIGME devices. (A1)
    and (B1): 3D view of the diskoids. Blue: The surrounding ECM
    embedded with fluorescent particles. Green: GFP-labeled MDA-MB-231
    cells. The seeding geometry of the diskoids are outlined in
    white. (A2-A3) Top and side views of the hexagonal diskoid at day 0 and day
    5. (B2-B3), Top and side views of the triangular diskoid at day 0 and day
    5. Scale bars: 200 $\mu$m.}
  \label{fig4}
\end{figure}

Solid tumors may develop various shapes \textit{in vivo}, resulting in
a diverse range of interfacial geometry between the cells and the
ECM. DIGME allows us to easily control the geometry of diskoid. To
demonstrate the capability, we employed laser-micromachining to
fabricate stainless steel needles with hexagon and triangle
cross-sections. Using these needles as the mold, we have generated
hexagonal and triangular MDA-MB-231 diskoids in 3D collagen ECM
(Fig. \ref{fig4} A1 and B1). We find that the original shapes of the
diskoid (Fig. \ref{fig4} A2 and B2) largely determine the invasion
pattern after 5 days of incubation (Fig. \ref{fig4} A3 and
B3). Previously it was reported that the interfacial geometry
regulates the tumorigenicity by promoting cancer-stem cells
\cite{Kilian2016}. Using DIGME, we show that the geometric control can
be realized in truly 3D setups.

\begin{figure}[t]
\centering
  \includegraphics[width=0.75 \columnwidth]{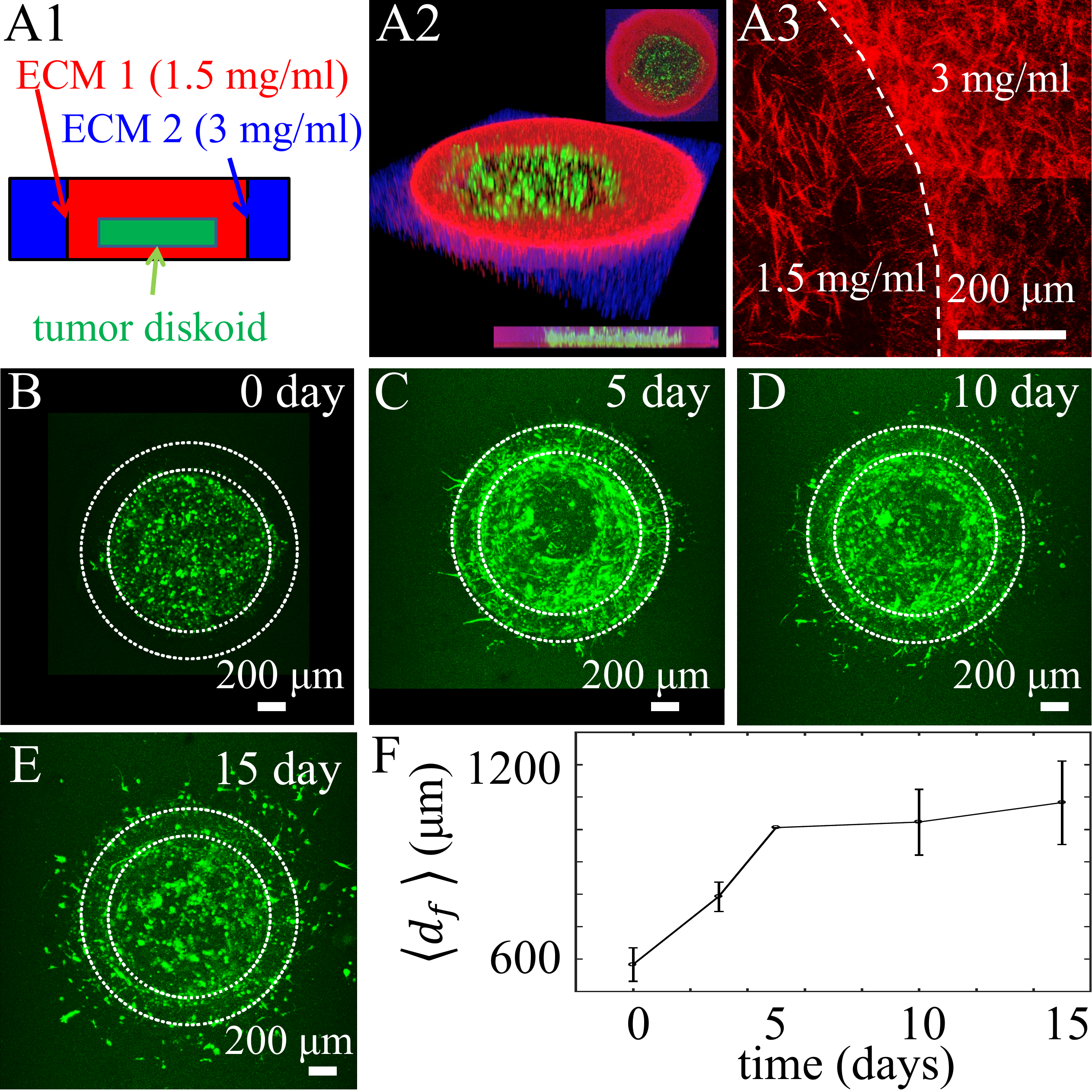}
  \caption{A two-layer DIGME device. A1: Schematics of the two-layer
  device. A circular MDA-MB-231 diskoid is confined by 1.5 mg/ml
  collagen matrix (ECM 1). ECM 1 is inside of 3 mg/ml collagen matrix
  (ECM 2). A2: 3D, top, and side views of the device. Green:
  MDA-MB-231 cells. Red: ECM 1 labeled with red fluorescent
  particles. Blue: ECM 2 labeled with far-red fluorescent
  particles. A3: confocal reflection image showing the collagen fibers
  at the interface of ECM 1
  and ECM 2. (B-E) Top views of the invading diskoid at day 0, 5, 10
  and 15. F: Invasion distance $d_f$ as a function of time.}
  \label{fig5}
\end{figure}
The extracellular space along the invasion path of a tumor is
spatially heterogeneous \cite{Chung2014}. Employing DIGME, we can
program the ECM heterogeneity and study its effect on the collective
cancer invasion. As a proof of concept, we have formed a MDA-MB-231
diskoid confined within two-layer ECM. This is done by sequentially
applying two circular molds of different diameters, and filling the
coaxial wells with different concentrations of collagen gels
(Fig. \ref{fig5}A1-A3). The inner layer, with collagen concentration
1.5 mg/ml is more porous compared to the outer layer, which has
collagen at concentration of 3 mg/ml (Fig. \ref{fig5}A3). We have
observed the invasion of the diskoid for over 15 days, and the top
views of the sample at day 0, 5, 10 and 15 are shown in
Fig. \ref{fig5}B-E. Within 5 days after initial seeding, invasion
front quickly reach at the interface of inner and outer layer of the
ECM (Fig. \ref{fig5}C). At day 5, most front cells polarize
tangentially at the interface, with a few leading cells polarize
radially and start to invade into the outer ECM layer. By quantifying
the invasion front profiles, we find that that invasion speed is
significantly reduced at day 5. Due to the change of cell orientation
as well as the ECM microstructure, the invasion speed reduces rather
abruptly at the interface between the two ECM layers.
\begin{figure}[t]
\centering
  \includegraphics[width=0.75 \columnwidth]{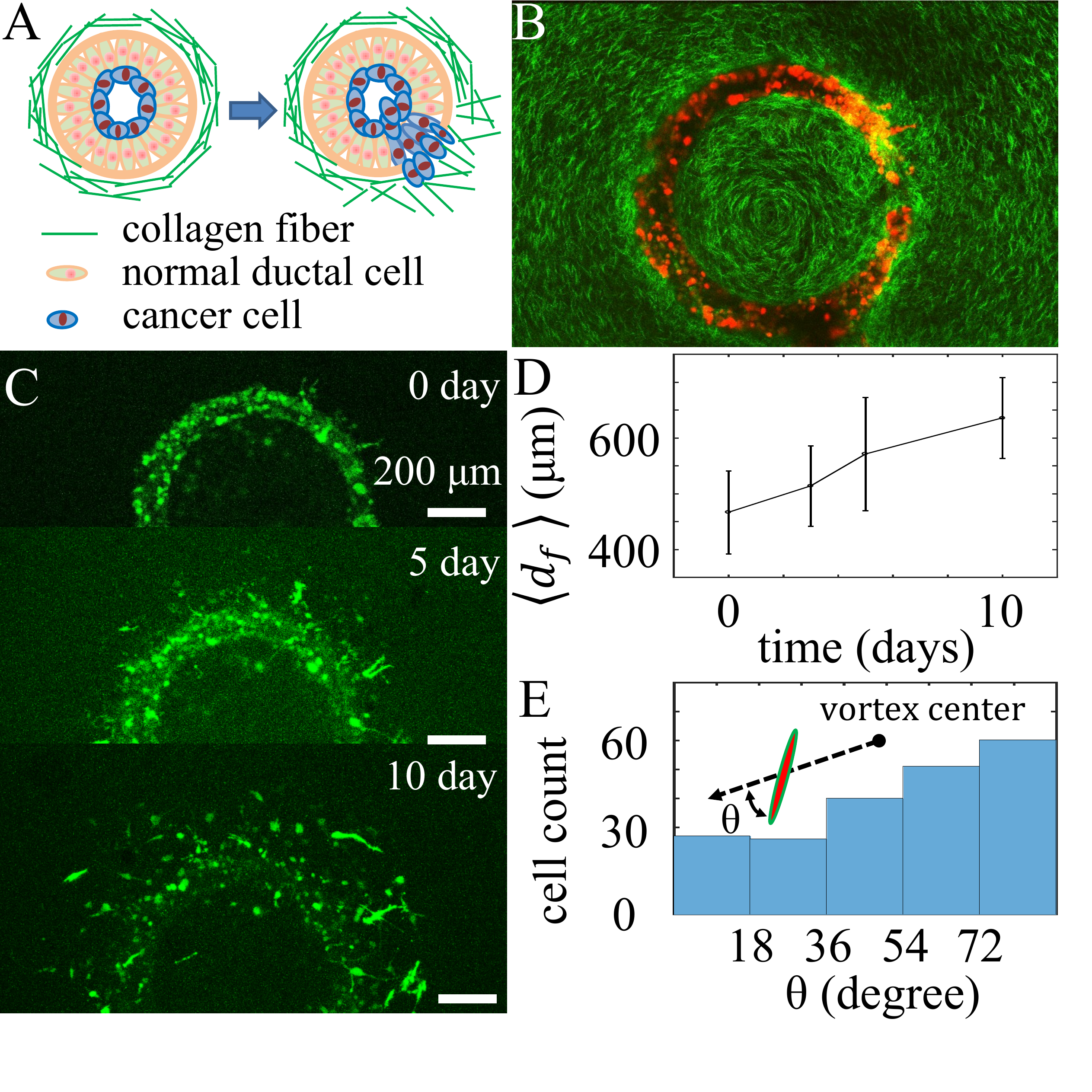}
  \caption{A ring diskoid in circularly aligned collagen matrix
    simulating a ductal carcinoma. A:
    schematics showing the invasion of a typical ductal
    carcinoma. B: A confocal slice showing the MDA-MB-231 ring diskoid
    (red) surrounded by circularly polarized collagen fibers
    (green). C: Top views of the sample at day 0, 5, and 10. D:
    Invasion distance $d_f$ as a function of time. E: Histogram of
    cell orientation $\theta$ at 10 days. $\theta$ is the
    angle between the cell long axis and the local radial direction measured
    from the seeding center of the diskoid $C_v$.}
  \label{fig6}
\end{figure}
Ductal carcinoma is the most common type of breast cancer. At the
early stage of ductal carcinoma, tumor cells are surrounded by
collagen matrix that are polarized long the tumor-stromal interface (Fig. \ref{fig6}A) \cite{Keely2006}. The orientation of
collagen fibers becomes disorganized even perpendicularly aligned during tumor
progression, correlating with the clinical outcome of cancer patients \cite{Keely2011}. Employing DIGME, we can control the orientation of collagen
fibers in the surrounding matrix, mimicking different stages of ductal carcinoma. As an example, we mount a 150 $\mu$m diameter
needle approximately 300 $\mu$m off the rotational axis of DIGME. While the 1.5 mg/ml
collagen gel is forming, we continuously rotate the needle at 1 Hz for
5 minutes. The microscopic flow driven by the needle aligns the collagen fibers, and the
fiber orientation is subsequently locked by the gelation process \cite{Kaufman2007}. At
the same time, the
rotating needle carves a ring in the collagen gel, which we fill with
MDA-MB-231 cells mixed in the host ECM (Fig. \ref{fig6}B). The host
ECM consists of 1.5 mg/ml collagen matrix that is randomly
oriented. We have imaged the invasion process of the diskoid for 10
days, and find that the circularly polarized collagen fibers strongly
impact the motility and morphology of MDA-MB-231 cells. The front
invasion speed (Fig. \ref{fig6}D) is noticeably slower compared with
the diskoid in randomly oriented matrix (Fig. \ref{fig2}B), and that a
large fraction of cells are oriented tangentially along the collagen
fibers (Fig. \ref{fig6}E). These results are consistent with the
contact guidance observed for single cells in
both 2D and 3D cultures \cite{Levchenko2012}.

\section*{Discussion}
We have described DIGME as a low-cost, easy-to-implement strategy to
engineer the geometric microenvironment of tumor organoids. The
shape-programmable organoid - diskoid, preserves the native cell-cell
contacts in 3D ECM, and allow us to study the single cell dissemination
and cohesive progression during collective cancer invasion. 

For a thin circular diskoid in isotropic homogeneous ECM, we have
shown that the collective invasion and morphological evolution of
MDA-MB-231 cells (Fig. \ref{fig1} to Fig. \ref{fig3}) follow the
similar patterns observed in the middle plane of tumor spheroids
\cite{Wirtz2015}. Tumor spheroids are often too dense to image through
directly, and requires destructive pre-imaging preparations such as
cryo-section. DIGME, on the other hand, provides an alternative model
allowing continuous, long-term imaging at the single cell level.

We find that the invasion profile correlates with the seeding
geometry, as shown in the hexagonal and triangular diskoids
(Fig. \ref{fig4}). It has been proposed that physical forces generated
by the cellular traction propagate over the ECM, and coordinate the 3D
collective cancer invasion \cite{Shenoy2014,Wirtz2015,Jiao2016}. By
controlling the shape of diskoids as well as the ECM microstructure,
we can tune the stress distribution in the ECM. Therefore DIGME
provides an ideal experimental system to understand the mechanical
mechanisms coordinating the self-organized collective cancer invasion.

We find that 3D collective cancer invasion is regulated by the spatial
heterogeneity, as well as the microscopic anisotropy of the ECM. A
progressing tumor encounters dramatically varying microenvironment, or
microniches \cite{Sauvage2013}. By employing DIGME, we can further extend the examples
demonstrated in Fig. \ref{fig5} and Fig. \ref{fig6} to generate
complex microniches in the ECM. For instance, epithelial cells and
fibroblast cells can be embedded in different layers of ECM. Also the
level of ECM fiber alignment can be controlled by varying the
rotational protocol that drives the DIGME mold. Such protocols,
including changing the rotational speed, or implementing bidirectional
rotation, can be easily realized with a proper choice of the rotary
motor in DIGME.

In the above, we have used breast cancer cell line MDA-MB-231 cells to
demonstrate the capabilities of DIGME. It is expected that DIGME
methods equally apply to any cells compatible with 3D culture, such as
fibroblast cells, endothelial cells, stem cells and neuron
cells. Similarly, type I collagen ECM can be replaced by other forms
of ECM in DIGME, including tissue-derived ECM like matrigel, and
synthetic ECM like peptide gel. With these extensions, DIGME is not
only useful to probe the collective invasion of tumors, but also
allows one to study 3D multicellular dynamics in wound healing,
angiogenesis, development, and tissue remodeling.

We notice that the current form of DIGME has several
limitations. First of all, DIGME only control the cross-sectional
shape of the diskoid, rather than the full 3D geometry of the tumor
organoid. Although variants of DIGME is possible, for instance by
using a cone-shaped mold, fully 3D patterning may require
incorporating other techniques such as directed self-assembly
\cite{Nelson2014}. Second, metal mold fabrication has a typical
tolerance of tens of micrometers, or the size of a cell. In order to
control the diskoid shape down to subcellular accuracy, one may
incorporate other micro-fabrication techniques, such as polymer laser
micromachining or deep reactive-ion etching
\cite{Sun2011landscape}. These alternative fabrication methods have
micrometer $\mu$m resolution, but may require surface treatment to
ensure low-binding affinity to collagen. Finally, the mechanical-based
DIGME method requires one to two hours to prepare each sample. To
improve the throughput, one may take advantage of the low-cost and
simple operation of DIGME and implement automated parallel processing.

\section*{Author Contribution}
B. S. designed the research, A. A. performed experiments, A. A. and
B. S. analyzed data and wrote the paper.
\section*{Acknowledgements}
A. A. is supported by a scholarship from the Culture Mission of the Royal Embassy of Saudi Arabia (SACM). B. S. thanks Oregon State University for startup support.

\section*{References}

\bibliography{DIGMEbibfile}

\end{document}